\documentclass{eas}
\usepackage{graphicx}
\usepackage{natbib}
\begin{document}

\title{Star Formation during Galaxy Formation}
\author{Bruce G. Elmegreen}\address{IBM T. J. Watson Research Center, 1101 Kitchawan Road, Yorktown
Heights, New York 10598 USA, bge@us.ibm.com}
\begin{abstract}
Young galaxies are clumpy, gas-rich, and highly turbulent. Star
formation appears to occur by gravitational instabilities in galactic
disks. The high dispersion makes the clumps massive and the disks
thick. The star formation rate should be comparable to the gas
accretion rate of the whole galaxy, because star formation is usually
rapid and the gas would be depleted quickly otherwise. The empirical
laws for star formation found locally hold at redshifts around 2,
although the molecular gas consumption time appears to be smaller, and
mergers appear to form stars with a slightly higher efficiency than the
majority of disk galaxies.
\end{abstract}
\maketitle

\section{Introduction}

In the first four lectures of this series, we reviewed star formation
in local galaxies. Recall that the star formation rate is proportional
to the CO emission, from which we concluded that star formation occurs
only in molecular gas and that the consumption rate from molecules to
gas is constant. This derivation assumed a fixed CO to H$_2$ conversion
rate to get the molecular gas mass, a fixed IMF, uniform grain
properties, and certain extinction corrections to get the star
formation rate. In addition, the molecular fraction scales almost
linearly with pressure, and the pressure depends on the mass column
densities, $\Sigma_{\rm gas}$ and $\Sigma_{\rm stars}$, and the
velocity dispersions, $\sigma_{\rm gas}$ and $\sigma_{\rm stars}$. We
also saw that spiral waves promote star formation in the arms, or
organize the star formation, but do not affect the average rate much.
The same is apparently true for star formation in shells, bright rims
and pillars, which trigger star formation in these regions, i.e.,
organize where it happens, without changing the global average rate
much. Star formation seems saturated in inner disks, so the detailed
mechanisms of cloud formation do not appear to matter.

In addition, we saw that stars form in hierarchical patterns with star
complexes, OB associations, clusters, and so on, because of turbulence
compression and self-gravity. As a result, there are power-law mass
functions for clouds, clusters, and stars, and there are space-time
correlations for clusters. There are probably similar space-time
correlations for young stars which are not observed yet.

We would like to discuss here what changes for young galaxies at high
redshift. At first, we expect high redshift galaxies to look like
normal galaxies viewed in the restframe ultraviolet. They would look
dimmer because of cosmological surface brightness dimming, and the star
formation regions would be blurred out because of poor spatial
resolution. But still, we might expect to see the uv restframe versions
of normal galactic features, i.e., exponential disks, spiral arms,
bars, lots of small star-forming regions, and a general diversity in
the relative prominence of disks and spheroids (i.e., the Hubble
types).

\cite{barden08} made model images of redshifted SDSS galaxies to
$z=0.15, 0.5,$ and 1, and even increased the intrinsic brightness for
the $z=1$ images. The result was a significant loss of faint
structures, including the outer disk and the faint star-forming
regions. \cite{overzier10} redshifted ``Lyman Break Analogs'' to
$z=2,3,$ and 4. They found that small clumps blend together and faint
peripheral tidal features disappear. Petty et al. (2009) looked at the
standard structural measures: the Gini coefficient, M20 (central
concentration), and the Sersic index for redshifted local galaxies.
They found that the model galaxies were smoother (lower Gini) and more
centrally concentrated (lower M20) than their local counterparts. These
studies reinforce our notion that high redshift galaxies should look
somewhat smooth and centrally concentrated if they are at all like
local galaxies.

\begin{figure}[b]
\begin{center}
\includegraphics[width=3.in]{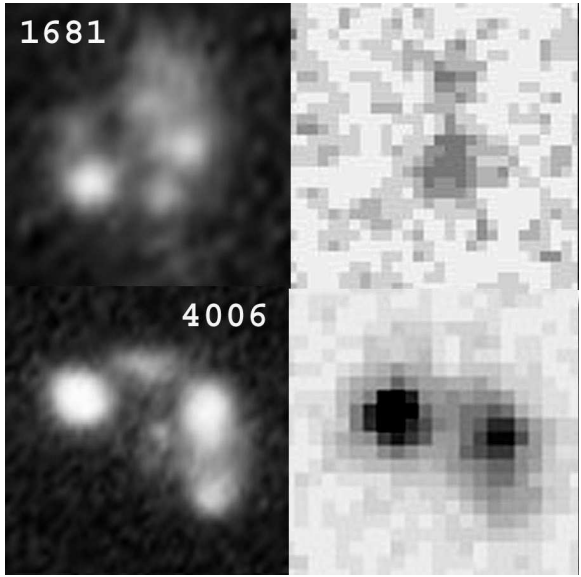}
\caption{Two examples of clump cluster galaxies from the HST UDF with
color ACS camera images on the left and NICMOS images on the right
\citep[from][]{e09a}. These galaxies are characteristic of this class,
having several large clumps of star formation and no obvious interclump
disk.}\label{f1}
\end{center}
\end{figure}

In fact, when deep high resolution images of the sky were taken,
particularly by the Hubble Space Telescope (HST), disk galaxies did not
look anything like these expectations from local galaxies. Beyond
$z\sim2$, galaxies are mostly irregular, asymmetric, and clumpy
\citep{vdb96,ab96,conselice05}. In particular, there is a class of
galaxies that is almost entirely clumpy, with nearly half of the light
in several big star-forming clumps and no obvious underlying
exponential disk \citep{ee05}. Figure \ref{f1} shows two examples of
clumpy disks, with UDF catalog numbers indicated. On the left are
SkyWalker\footnote{designed by K. Jahnke and S.F. S\'anchez, AIP 2004}
images using the ACS camera and on the right are NICMOS images in the
near-infrared with $3\times$ lower resolution. The galaxies contain
several large star-forming clumps with no central concentration from a
bulge or exponential disk.

In a catalog of galaxy morphologies in the HST Ultra Deep Field,
considering only galaxies larger than 10 pixels so their internal
structure can be observed, \cite{e05a} recognized 6 basic types: Chain
galaxies (121 examples), Clump Clusters (192), Double (134), Tadpole
(114), Spiral (313), and Elliptical (129). Only the spirals and
ellipticals resemble local galaxies, and even then the spirals tend to
have bigger star-forming regions than local spirals and the ellipticals
are clumpy as well \citep{e05b}. Photometric redshifts of these
galaxies \citep{e07a} suggested that all of the clumpy types extend out
to at least $z\sim5$ with extreme starburst spectral energy
distributions. The spirals and ellipticals end at about $z\sim1.5$,
which could be because either their number density drops, or they are
too faint in the restframe uv to see at higher redshifts.

\section{What are the Clumpy Types?}

Chain galaxies were originally identified by \cite{cowie95} using
ground-based images. They are linear objects with several giant clumps
along their length. There is often no central red clump, and no
exponential profile as in a modern edge-on spiral galaxy. There are
also many oval-shaped clumpy galaxies that resemble chains in having
the same numbers, magnitudes, and colors for the clumps. More
important, the relative numbers of these systems, chains versus clumpy
galaxies, is consistent with the chains being edge-on clump clusters
\citep{eeh04a,ee05}. Thus we have a new morphological type of galaxy, a
thoroughly clumpy disk viewed in random orientations, that occurs
primarily at high redshift.

Chains and clump clusters are so common that all modern spirals could
have gone through this phase at $z> 1$.  Essentially all observed disk
systems are very clumpy at $z>2$. The comoving space density of chains
and clump clusters larger than 10 pixels in the UDF is $\sim4\times
10^{-3}$ Mpc$^{-3}$ for $z<1$, decreasing to $\sim1\times10^{-3}$
Mpc$^{-3}$ out to $z\sim3$ or more. For spirals larger than 10 pixels
in the UDF, the space density is $4\times10^{-3}$ Mpc$^{-3}$ for $z<1$,
but decreasing faster with $z$ than the clumpy types, perhaps, in part,
because spirals become too red to see.  Considering also that the
clumpy phase is probably shorter lived than the spiral phase, the
prevalence of clumpy disks at high redshift seems clear.

Most clumps are not a bandshifting artifact of rest-UV normal star
formation. Clumpy and spiral types are both present at low redshift. In
GOODS, there are four basic types of disk galaxies: density wave
spirals and flocculent spirals resembling today's galaxies, clumpy
galaxies with a red disk between the clumps, and clumpy galaxies
without any evident disk between the clumps. All four types span the
same range of redshifts up to $z\sim1$. There are clump clusters even
at $z\sim0.2$. This is such a low redshift that the observed V band in
GOODS corresponds to a restframe passband of B band. Local spirals do
not look like clump clusters in B band, so the clumpies are
intrinsically different.

\section{Mergers}

Highly irregular massive disk galaxies in the local universe are
usually mergers or interacting systems. We don't know if this is also
true at high redshift. In the GOODS sample, there are clump clusters
and chains at low redshift that look the same as those at high redshift
in the UDF. But also in GOODS there are many examples of mergers and
interactions that look like their local counterparts
\citep{ee06a,e07b}. Thus normal mergers and interactions show up just
fine in GOODS, and clump clusters are different. Clump clusters usually
have no tidal features, for example, and they do not typically have
double red nuclei from formerly separate galaxies.

Mergers are also not required to make a galaxy lopsided. Internal
processes can do that too. \cite{bournaud08} observed a lopsided clump
cluster in the UDF with Sinfoni. They found that it has a smooth
rotation profile and metallicity gradient, so it does not look like a
chaotic merger or have a double rotation curve.  A simulation of this
system reproduced the lopsided shape very well if the initial disk and
halo were offset from each other a little.  This offset seems
reasonable if very young galaxies undergo rapid accretion from a
cosmological inflow; they should often have their disk center-of-mass
at a slightly different position than their halo center-of-mass.  The
disk mass in this model was $6\times10^{10}\;M_\odot$, with half of the
disk mass in gas.

The final piece of evidence that chains are edge-on clump clusters is
that the clumps in chain galaxies are highly confined to the midplanes.
Their resolution-corrected rms deviation from the midplane of the
chains is less than 100 pc \citep{ee06b}. This requires {\it in situ}
clump formation, not extra-galactic clump accretion. Also some chains
are curved and not straight, and the clumps in them follow the
curvature too, without significant deviations \citep{ee06a}. These may
be interacting edge-on clumpy galaxies, but still the clumps formed in
them and are not separate merger remnants.

\section{Clump Cluster Properties}

Clumpy young galaxies, whether somewhat face-on and called ``clump
clusters,'' or edge-on and called chains, have properties that are
consistent with their youth, and also show variations that are
consistent with their gradual evolution into modern disks.

Their youthful appearance is reinforced by the observation that they
are highly molecular \citep{tacc10,daddi08, daddi10a} and highly
turbulent \citep[][see below]{forster09,law09}. Presumably the
turbulence is a result of energy gained from intergalactic accretion
\citep{eb10} and gravitational instabilities in the disk
\citep{bournaud09}. Many young galaxies have a ratio of rotation speed
to twice the dispersion speed that is less than unity. \cite{forster09}
consider that when this ratio is less than 0.4, the disks are
dispersion-dominated. Their galaxies have stellar masses in the range
from $10^{10}\;M_\odot$ to $10^{11}\;M_\odot$ and dynamical masses that
are 3 to 5 times larger, on average.

\cite{tacc10} made CO (3-2) maps of several clump clusters and found
that typical clumps in these galaxies have $5\times10^9\;M_\odot$ of
H$_2$ with radii $< 1$-2 kpc, $\Sigma_{\rm H2}=300-700\;M_\odot$
pc$^{-2}$, and $\sigma\sim19$ km s$^{-1}$.  They also derived a high
gas fraction in the disks.  \cite{daddi10a} observed CO in 6 galaxies
at $z\sim1.5$, finding rotation in some cases and a generally high gas
fraction. The timescales for gas consumption, stellar build-up, and
galactic dynamics were all comparable in the Daddi et al. study, which
implies that the galaxies are very young. Daddi et al. also found that
the efficiency of star formation is about the same as in normal
galaxies today.

The clump stellar masses in clump clusters are $\sim100$ times larger
than star complex masses in modern spiral galaxies of similar
luminosities \citep{e09b}. Bulges or bulge-like objects in clump
clusters and chains are sometimes observed, and they are more like the
clumps in terms of mass and age than are the bulges in spiral galaxies
\citep{e09a}. The interclump surface density and age relative to the
clump surface density and age also show variations among different
clump clusters \citep{e09b}. All of these variations suggest an
evolution from highly clumpy, bulge-free galaxies to smooth spiral
galaxies with bulges.

There are essentially no barred clump cluster galaxies. Even if bars
were present, they could hardly be recognized in such irregular disks.
Bars appear only when the galaxies calm down and develop exponential
disks and central concentrations or bulges.  Still, there are elongated
clumps in some clump clusters, suggesting protobars \citep{eeh04b}. If
these objects really turn into bars, then this suggests bar formation
can be a gas-rich process, including significant energy dissipation,
and not a pure stellar process as in standard numerical models.

\section{Working Model}

The most likely model for the origin of clumps in clumpy galaxies is
that they form by gravitational instabilities in rapidly assembled
disks. The clumps are confined to within 100 pc of the mean disk, they
are young star-forming regions (not diverse merged galaxies), the clump
masses are $10^7\;M_\odot$ -- $10^8\;M_\odot$, sometimes
$10^9\;M_\odot$, and these masses appear to be the ISM Jeans masses
with the measured turbulent speeds and gas column densities. For
example, $M_{\rm Jeans}\sim \sigma^4/G^2\Sigma\sim10^8\;M_\odot$ if
$\sigma\sim30$--50 km s$^{-1}$ and $\Sigma_{\rm gas}\sim100\;M_\odot$
pc$^{-2}$. These dispersions are consistent with observed HII
dispersions \citep{forster06,forster09,weiner06,genzel06,genzel08,
puech07,law09} and this column density is typical for the inner disk
regions of spiral galaxies today. It is also comparable to what
\cite{tacc10} observed directly using CO emission.

There are many consequences of having such large clumps in a galaxy
disk \citep{noguchi99,immeli04a,immeli04b,bournaud07}. They contribute
strongly to the total disk potential, so they interact gravitationally,
experience strong dynamical friction, and lose angular momentum to the
outer disk. This all causes them to migrate rather quickly to the disk
center where they contribute to a growing bulge \citep{e08a}. Star
formation in the center can get triggered by their merger too, and this
adds to the bulge. At the same time, their disruption in the disk
causes it to smooth out, and this, combined with their angular momentum
transfer, gives the disk an exponential radial profile
\citep{bournaud07}. All of this disk evolution can happen within 0.5-1
Gyr.

Stirring from the clumps also thickens the disk and this probably
produces the thick disk component of today's spiral galaxies
\citep{bournaud09}.  Thick disks can also form by minor mergers, both
through the stirring of existing disk stars and the dispersal of the
merger remnants \citep{quinn93,walker96}. However, thick disks formed
in this way flare out at the edge, and real thick disks do not seem to
do this \citep{yoachim06,be09}. Stirring by internal processes
automatically makes a thick disk with an approximately constant scale
height, because the stirring force from clump gravity is proportional
to the disk restoring force from gravity. In the case of a merger, the
stirring force is proportional to the companion galaxy mass and
independent of disk restoring force, so the disk is dispersed much
further in the outer regions than the inner regions \citep{be09}.

There is also a possible connection with nuclear black holes if the
dense clusters that are likely to be present in the cores of the
individual disk clumps form intermediate mass black holes by stellar
coalescence, as proposed for dense clusters by \cite{ebi01},
\cite{port02}, and others. If the clumps form black holes in this way,
then these black holes will migrate into the disk center along with the
clumps, and possibly merge to make a massive nuclear black hole.
Simulations of this process obtain a correlation between the black hole
mass and the bulge velocity dispersion that is similar to what is
observed \citep{e08b}. If the clumps make globular clusters too
\citep{shapiro10}, then the correlation between globular cluster number
and central black hole mass \citep{burkert10} might be explained in the
same way.

\section{Stream-fed Disks}

Galaxy accretion by cold gas streams is a way to feed gas into the
disks fast enough to produce wild instabilities and clump formation.
This can all happen without galaxy mergers, except for some small
galaxy-like pieces that come in with the cold flows. The recognition of
cold flows is a major change in thinking about how galaxies form
\citep{murali02,birn03,semelin05,db06,ocvirk08,dekel09a,dekel09b,
ager09,ker05,ker09,brooks09}. Hierarchical build-up models in the cold
dark matter scenario may not apply to baryons as much as they apply to
cold dark matter itself. The baryons may enter a galaxy in the form of
cold flows, rather than minor and major mergers of component galaxies,
each with their own dark matter halo.

\cite{ceverino10} modeled cold and hot flows with a disk galaxy forming
in the center. The model is appropriate for a redshift of $z=2.3$. They
follow the formation and evolution of individual clumps in the disk
gas, showing how the accretion quickly makes an unstable gas disk,
which forms giant clumps that migrate to the center.

\section{Local analogs of clumpy galaxies}

Clumpy galaxies do not look like local galaxies even when the local
galaxies are modified to appear as they would at high redshift. FUV
images of local galaxies contain too many star-forming clumps, and they
are also more centrally concentrated than clump clusters. Spiral and
barred structure in local galaxies would still show up at high redshift
too (using the HST ACS camera, for example), if the disk is not too
faint to see.

UDF clump clusters have bigger and fewer clumps than local galaxies,
even in the restframe uv, they have no symmetry or central
concentration, and they are much brighter in restframe magnitudes.
Typical clump clusters have surface brightnesses that are more than 10
times larger than the surface brightness of, for example, M101, which
is a locally bright galaxy with lots of giant star-formation clumps
(although the M101 clumps are still small by high redshift standards).

We may wonder if local flocculent spiral galaxies are a better match to
high-$z$ galaxies because local flocculents get most of their structure
from gravitational instabilities in the gas and there are no prominent
spiral waves in the old stellar disk. Two redshifted versions of the
flocculent galaxy NGC 7793 were shown in \cite{e09b} and compared to
GEMS galaxies. The local and distant galaxies do not look similar at
all. In general, local galaxies are too smooth and too centrally
concentrated compared to clump clusters.

On the other hand, a local dwarf Irregular galaxy is a good match to a
clump cluster, although the clump clusters are much more massive
\citep{e09b}. Clump clusters resemble local dwarf irregulars because
both have high gas fractions, both have big complexes relative to the
galaxy size, both have relatively thick disks, and both have high
velocity dispersions relative to the rotation speed. Recall that
$L_{\rm Jeans}/{\rm Galaxy Size}\sim H_{\rm disk}/{\rm Galaxy
Size}\sim(\sigma/V)^2$. That is, the clump size from gravitational
instabilities is comparable to the galactic scale height, and the ratio
of these lengths to the galaxy size is the square of the ratio of the
velocity dispersion to the rotation speed. Thus big complexes, thick
disks, and high dispersions (relative to galaxy size and rotation
speed) all go together regardless of the galaxy mass.

Both local dwarf irregulars and clump clusters are irregular because
they have a relatively high gas mass and a high $\sigma/V$. Both are
also relatively young in terms of the number of rotations they have
lived and in terms of the relative gas abundance. The resemblance
between clump clusters and local dwarfs is another example of down
sizing: small galaxies today (dwarf irregulars) are doing what big
galaxies (clump clusters) did at $z\sim2$ \citep{e09b}.

There are other local galaxies that resemble clump clusters too, but
they have about the same stellar mass as clump clusters, i.e., they are
large galaxies. These local analogues are extremely rare, however.
\cite{casini76a,casini76b} and \cite{maehara88} discovered local
``clumpy irregular galaxies'' of normal size. Examples are Markarian
296, 325, 7, 8 (which are ultraviolet galaxies), and Kiso UV excess
galaxies 1618+378, 1624+404, 1626+413, and Mrk 297. Maehara et al.
(1988) determined galactic distances of 60 to 120 Mpc, clump sizes of
$\sim2^{\prime\prime}$ (corresponding to 1 kpc), and clump absolute
magnitudes of $M_{\rm B}\sim-11$ to $-16$ mag (corresponding to
$\sim10^6\;L_\odot$ to $10^8\;L_\odot$).

\cite{garland07} studied "Luminous Compact Blue Galaxies." These are
small, high luminosity, high surface brightness galaxies with a blue
color. They are also gas-rich (CO, HI), like high-z galaxies, and
rotating with distorted velocities, as if they are interacting or
lopsided. \cite{overzier08,overzier09,overzier10} studied Lyman Break
Analogs \citep{heckman05}. These are super-compact uv-luminous
galaxies. They are GALEX objects with $L_{\rm FUV}>10^{10.3}\;L_\odot$
and intensities $I_{\rm FUV}>10^9\;L_\odot$ kpc$^{-2}$ at redshifts
$z<0.3$. They are also very rare ($\sim10^{-6}$ Mpc$^{-3}$).

\section{What should a Model of Star Formation be for High Redshift
Galaxies?}

Young galaxies look like their whole disk is out of equilibrium. In
general terms, the star formation rate is expected to equal the
accretion rate. This implies that if simple laws like the KS relation
or the Bigiel-Leroy relation apply, then they fix the gas column
density or molecular column density for a given star formation rate,
not the other way around.  Maybe the molecular abundance still depends
on pressure and the radiation field, and maybe stars still form only in
molecular gas, but if the star formation rate is pinned to the
accretion rate, then these local relations are not useful and in
predicting the star formation rate. Perhaps the growth rate of GMCs
equals the accretion rate by a whole galaxy. This would seem to be
necessary to maintain a steady state. In a broad sense, this situation
is like the dynamical triggering models discussed in these Lectures
earlier, i.e., the spiral-wave or shell-like organization of gas into
star-forming regions. Instead of spiral waves and shells collecting
matter on a kpc scale, whole galaxies are collecting matter on a 10 kpc
scale.

\section{Comparison of Star Formation Models}

\cite{choi10} compared three models for star formation in cosmological
simulations, the \cite{springel03} model, the \cite{blitz06} model, and
the \cite{schaye08} model. These are instructive to review here so that
the diversity of analytical models can be noted.

In the \cite{springel03} model, there is cooling gas with a star
formation rate \begin{equation} d\rho_*/dt=(1-\beta)\rho_{\rm c}/t_{\rm
SFR},\end{equation} where $\beta$ equals the supernova gas return
fraction, $\rho_{\rm c}$ equals the density of cool clouds, $t_{\rm
SFR}=t_0^*(\rho/\rho_{\rm th})^{-1/2}$ is the dynamical time, where
$t_0^*=2.1$ Gyr gives the local KS law. They also assumed $\Sigma_{\rm
SFR}=0$ if $\Sigma<\Sigma_{\rm th}$  for threshold column density
$\Sigma_{\rm th}$, and $\Sigma_{\rm SFR}= A(\Sigma_{\rm
gas}/1\;M_\odot\;{\rm pc}^2)^n$ for $A=2.5\pm0.7\;M_\odot$ yr$^{-1}$
kpc$^{-2}$, $n=1.4$, and $\Sigma_{\rm th}=10\;M_\odot$ pc$^{-2}$. This
equation for star formation rate is combined with another equation for
the rate of change of the cool cloud density, \begin{equation}
d\rho_{\rm c}/dt=C\beta\rho_{\rm c}/t^*,\end{equation} where
$C=C_0(\rho/\rho_{\rm th})^{-4/5}$. This assumes that supernovae
evaporate and form cold clouds as in \cite{mo77}.

In the \cite{blitz06} model, the molecular fraction is given by
$\rho_{\rm H2}/\rho_{\rm HI}=(P_{\rm ext}/P_0)^{0.92}$, where
$P_0=4.3\times10^4$ k$_{\rm B}$ K cm$^{-3}$.  Then for a star formation
rate, they assume \begin{equation} d\rho_{*}/dt= (\rho_{\rm gas}/{\rm
Gyr})/[1+(P+P_0/P_{\rm ext})^{0.92}],\end{equation} which assumes
$\Sigma\propto\rho$ (a constant scale height). This star formation rate
was applied only when $P_{\rm ext}<P_0$. For $P_{\rm ext}>P_0$,
\cite{choi10} used the Springel \& Hernquist law (i.e., the Kennicutt
$n=1.4$ law).

The third model was that of \cite{schaye08}. These authors solved for
the scale height using $\Sigma_{\rm gas}=\rho_{\rm gas}L_{\rm Jeans}=
(\gamma f_{\rm g}P_{\rm tot}/G)^{1/2}$, where $\gamma$ is the adiabatic
index: $P_{\rm tot}\sim\rho_{\rm gas}^\gamma$ ($P_{\rm tot}$ and
$\rho_{\rm tot}$ include stars), $f_{\rm g}$ equals the gas mass
fraction, and $f_{\rm th}$ equals the thermal pressure fraction
($P=f_{\rm th}P_{\rm tot}$). They assumed $f_{\rm g}=f_{\rm th}$, so if
$\Sigma_{\rm SFR}=A\Sigma_{\rm gas}^n=\Sigma_{\rm gas}/t$ which means
$t=\Sigma_{\rm gas}^{1-n}/A$, then $t_{\rm SFR}=A^{-1}(M\odot\;{\rm
pc}^2)^n(\gamma P/G)^{(1-n)/2}$, and finally $d\rho_{*}/dt=\rho_{\rm
gas}/t_{\rm SFR}$. Note that in this model, $\Sigma_{\rm SFR}\propto
\Sigma_{\rm gas}P^{0.2}$. They also assumed a threshold density,
$\rho_{\rm th}=\Sigma_{\rm th}/L_{\rm Jeans}$, so $\rho_{\rm
th}=\Sigma_{\rm th}^2G/c_{\rm s}^2f_{\rm g}$ for $c_{\rm s}=1.8$ km
s$^{-1}$ (500K gas), and they assumed $P=K\rho^{4/3}$.

\cite{choi10} note that the Springel \& Hernquist model forms too many
stars at low $\Sigma_{\rm gas}$ and this causes it to form stars too
early in the Universe. The other models have a pressure dependence for
star formation which gives an acceptably low rate in low pressure
regions.

\cite{genzel10} reviewed the star formation and CO data for high-z
galaxies in the context of  the ``main-sequence line'' for star
formation:

\begin{equation} SFR (M_\odot \;{\rm yr}^{-1})=
150(M_*/10^{11}\;M_\odot)^{0.8}([1+z]/3.2)^{2.7}\end{equation}
\citep{bouche10,noeske07,daddi07}.

Genzel et al. noted that the gas depletion time depends weakly on $z$.
It is $\sim0.5$ Gyr at $z>1$, and $1.7$ Gyr at $z=0$, while mergers
have $2.5-7.5\times$ shorter depletion times than non-mergers. At
$z>1$, the depletion time is comparable to the stellar age, and it is
always shorter than the Hubble time. This means there is a continuous
need for gas replenishment in galaxy disks.

Genzel et al. also found that the molecular star formation-column
density relation is in agreement with \cite{bigiel08}. There is no
steepening at $\Sigma_{\rm gas}>100\; M_\odot$ pc$^{-2}$ like there
appears to be in local starbursts. In general, they find no variation
in the empirical star formation law with redshift.

The dynamical version of the \cite{ken98} relation was examined by
Genzel et al. too.  The dynamical relation says that the star formation
rate is proportional to the gas column density divided by the local
orbit time.  Even in this form, mergers were found to be more efficient
at star formation than non-mergers. The star formation efficiency per
unit dynamical time was about 1.7\%.

These considerations led Genzel et al. to a fundamental plane of star
formation, in which the total galactic star formation rate depends only
on the dynamical time and the total molecular mass:
\begin{equation}
\log\left({{SFR}\over{M_\odot/{\rm yr}}}\right) = -0.78\pm0.23
\log\left({{t_{\rm dyn}}\over{{\rm yr}}}\right) +1.37\pm0.16
\log\left({{M_{\rm mol}}\over{M_\odot}}\right)
-6.9\pm1.9,\end{equation} all with a standard deviation of 0.47 dex.
This is the same as \begin{equation} SFR=130\left({{M_{\rm
mol}}\over{10^{10}\;M_\odot}}\right)^{1.37} \left({{t_{\rm
dyn}}\over{100 {\rm Myr}}}\right)^{-0.78}\;M_\odot\;{\rm
yr}^{-1}.\end{equation}

\cite{genzel10} summarized the various star formation-density laws as
follows. The Kennicutt slope of $\sim1.4$ includes HI, whereas the
\cite{bigiel08} slope of $\sim1$ is just for CO (or H$_2$). Genzel et
al. redid the \cite{ken98} slope of 1.4 with just H$_2$ and got a slope
of $\sim1.33$. They redid their own relation for $\Sigma_{\rm SFR}$
versus $\Sigma_{\rm gas}$ including HI in addition to H$_2$ and found
that it increases the slope from 1.17 to 1.28. \cite{ken98} used the
same CO-H$_2$ conversion factor everywhere. When Genzel et al. redid
the \cite{ken98} data with a variable CO-H$_2$ conversion factor
including only H$_2$, the slope increased from 1.33 to 1.42. Thus the
inclusion of HI and the constant CO-H$_2$ conversion factor somewhat
cancel each other in the \cite{ken98} relation. \cite{genzel10} also
included new merger galaxies, which flatten the slope compared to that
in \cite{ken98}. Writing the star formation rate as $\Sigma_{\rm
mol}/t_{\rm dyn}$ works fairly well, including both mergers and normal
galaxies at all redshifts.

\cite{daddi10b} fitted Ultrahigh Luminosity Infrared Galaxies (ULIRGS)
and Submillimeter Wave Galaxies (SMGs) to the same star
formation-column density relation if the SF law is $\Sigma_{\rm
SFR}\sim\Sigma_{\rm gas}/t_{\rm dyn}$ ($t_{\rm dyn}$ is the rotation
time at the outer disk radius). They suggested that the global star
formation rate is proportional so  $\sim(M_{\rm  total\; gas} /t_{\rm
dyn})^{1.42}$. High SFR galaxies consume their gas faster than a
rotation time at the outer radius. This suggests that mergers are
involved, or some other rapid accretion leading to centralized star
formation.

\section{Summary}

Star formation at high redshift occurs in galaxies with high gas
fractions and high turbulent speeds. The morphology is a clumpy disk
without an exponential disk, bulge, or bar. There appear to be no exact
local analogues, but what comes close are the dwarf Irregular galaxies,
which are much lower in mass, along with very rare types called clumpy
irregular galaxies, luminous compact blue galaxies, Lyman Break
Analogs, etc.. All have a small number of giant star-forming regions,
presumably because they are all gas-rich and highly turbulent. The main
process of star formation everywhere seems to be gravitational
instabilities. The clump masses are large and the rates are large
because of the high turbulence and high gas mass fractions.  There is a
relation between the star formation rate and the galaxy mass versus
redshift, called the ``main sequence'' line. Cosmological models have
been made that fit this line fairly well. Generally, the $\Sigma_{\rm
SFR}$ versus $\Sigma_{\rm mol}$ relation is similar to that in local
galaxies, with a slight preference for a rate given by the galaxy
dynamical time rather than the pure gas dynamical time.  In galaxy
accretion models, the star formation rate quickly becomes equal to the
gas accretion rate. Presumably $\Sigma_{\rm mol}$ adjusts to
accommodate or enforce this equilibrium.


\end{document}